\date{}
\documentclass[10pt,aps,prl,twocolumn,superscriptaddress]{revtex4-1}
\usepackage{graphicx,psfrag,amsmath,amssymb,amsfonts,color}
\usepackage[T1]{fontenc}
\usepackage{natbib}
\usepackage[hidelinks]{hyperref}
\usepackage{upgreek}
\usepackage{color}
\usepackage{ulem}
\usepackage[nice]{nicefrac}

\newcommand{\ket}[1]{\ensuremath{\,|{#1}\rangle}}

\newcommand{\op}[1]{\hat{#1}}
\newcommand{\ve}[1]{\boldsymbol{#1}}

\newcommand{\beq}{\begin{equation}}
\newcommand{\eeq}{\end{equation}}
\newcommand{\bea}{\begin{eqnarray}}
\newcommand{\eea}{\end{eqnarray}}
\newcommand{\rone}{r_\mathrm{R1}}
\newcommand{\rtwo}{r_\mathrm{R2}}
\newcommand{\rax}{r_\mathrm{ax}}

\begin{document}


\title{Coherent transfer of the transverse momentum of an optical vortex beam \\to the motion of a single trapped ion}



\author{Felix Stopp}
\affiliation{QUANTUM, Institut für Physik, Universität Mainz, Staudingerweg 7, 55128 Mainz, Germany.}
\author{Maurizio Verde}
\affiliation{QUANTUM, Institut für Physik, Universität Mainz, Staudingerweg 7, 55128 Mainz, Germany.}
\author{Milton Katz}
\affiliation{Departamento de Física, FCEyN, Universidad de Buenos Aires and IFIBA, CONICET, Pabellón 1, Ciudad Universitaria, 1428 Ciudad de Buenos Aires, Argentina.}
\author{Martin Drechsler}
\affiliation{Departamento de Física, FCEyN, Universidad de Buenos Aires and IFIBA, CONICET, Pabellón 1, Ciudad Universitaria, 1428 Ciudad de Buenos Aires, Argentina.}
\author{Christian T. Schmiegelow}
\email{schmiegelow@df.uba.ar}
\affiliation{Departamento de Física, FCEyN, Universidad de Buenos Aires and IFIBA, CONICET, Pabellón 1, Ciudad Universitaria, 1428 Ciudad de Buenos Aires, Argentina.}
\author{ Ferdinand Schmidt-Kaler}
\affiliation{QUANTUM, Institut für Physik, Universität Mainz, Staudingerweg 7, 55128 Mainz, Germany.}

\begin{abstract}
We demonstrate the excitation, using a structured light beam carrying orbital angular momentum, of the center of mass motion of a single atom in the transverse direction to the beam's propagation.
This interaction is achieved with a vortex beam carrying one unit of orbital angular momentum and one unit of spin/polarization angular momentum. Using a singly charged $^{40}$Ca$^+$ ion, cooled near the ground state of motion in the 3D harmonic potential of a Paul trap, we probe the narrow S$_{1/2}$ to D$_{5/2}$ transition near 729~nm on its motional sidebands to quantify the momentum transfer.  Exchange of quanta in the perpendicular direction to the beam's wave vector $\ve{k}$ is observed in case of the vortex shaped beam, in strong contrast to the absence of this spin-motion coupling for the case of a Gaussian beam. We characterize the coherent interaction by an effective transverse Lamb-Dicke factor $\eta^\mathrm{exp}_{\perp}=0.0062(5)$ which is in agreement with our theoretical prediction $\eta^\mathrm{theo}_{\perp}=0.0057(1)$. 


\end{abstract} 

\maketitle

Light can impart both momentum and angular momentum on massive particles. The angular momentum of a light beam is determined not only by its polarization but also, by its spatial structure. For example, vortex beams, such as Laguerre-Gaussian beams, have been shown to carry orbital angular momentum (OAM) associated to their spatial structure~\cite{allen1992orbital}.  Since then, the use of structured beams has become a commonly applied technology including their use for imaging, communications, optical manipulation mechanics~\cite{padgett2017orbital} as well as for control and manipulation of cold atomic gases~\cite{amico2021roadmap}. In 1995 H. Rubensztein-Dunlop and her team showed that a trapped micron-sized particle could set to rotate with the sense sign of the beam's singularity~\cite{he1995direct}.
Following these ideas into the quantum regime, here we show that the angular momentum coming from the beam's structure can coherently excite the center-of-mass motion of a trapped atomic ion. In particular, we show the ion can be coherently driven in the plane perpendicular to the impinging direction, in striking contrast to Gaussian beams, where transversal momentum exchange is not possible. 

Here, we use a single trapped ion which can be positioned with nanometer precision with respect to a structured beam~\cite{drechsler2021optical}. 
After laser cooling of the motion, the ion's center of mass dynamics can be adequately described as a 3-dimensional quantum harmonic oscillator~\cite{stenholm1986semiclassical}.
Then, on a narrow ($\sim$Hz wide) electric quadrupole transition, the internal electronic and external vibrational quantum states of the ion can be prepared, manipulated and read out via coherent driving in the resolved sideband regime~\cite{leibfried2003quantum}. 
These prerequisites allow for pushing the seminal experiments where light's orbital angular momentum can be transferred to the motion of a single trapped particle to the quantum regime. In a previous experiment, we showed that orbital angular momentum can be transferred to the motion of the valence electron of an ion~\cite{schmiegelow2016transfer}, while in this work we show that the beam's orbital angular momentum can coherently excite the atom's motion as a whole, i.e. to its center of mass.  

\begin{figure}[tp]
\centering\includegraphics[width=0.48\textwidth]{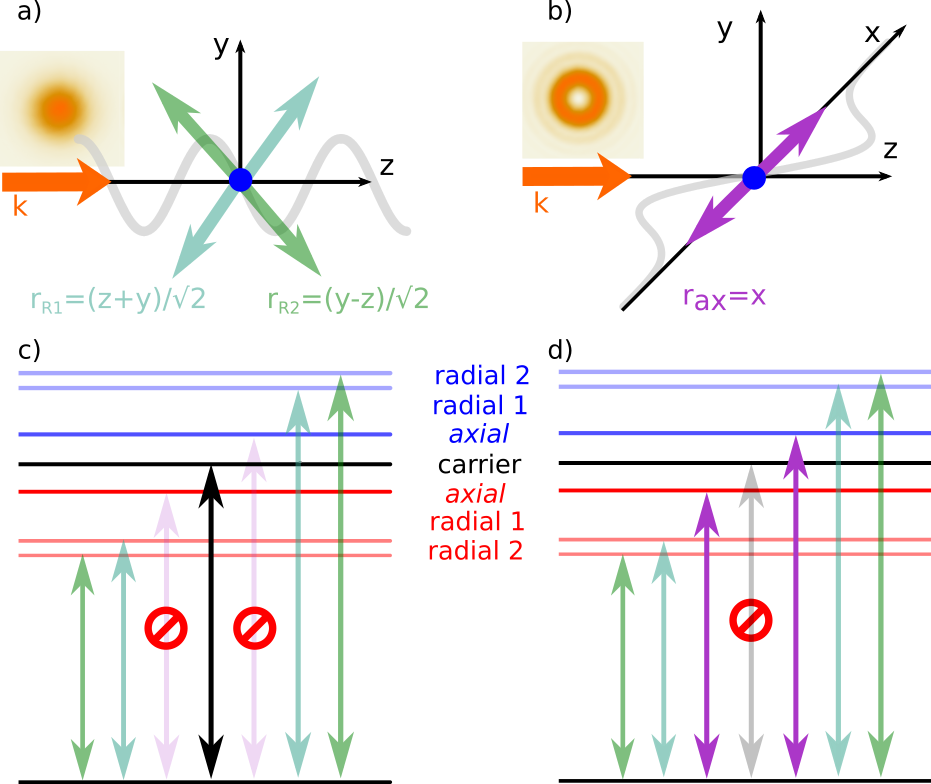}
\caption{Light field traveling along the $z$ direction for a) a Gaussian beam and b) a OAM-carrying vortex beam. The coordinate system of the beam is expressed by $x$, $y$, $z$, while an ion localized in the origin has the motional modes in $\rone$, $\rtwo$ and $\rax$ direction. c) $\varDelta m=\pm1$-transitions of a Gaussian beam with $\sigma=\pm1$: While carrier and radial sidebands are allowed, the axial sidebands transitions are forbidden, because of the orthogonal orientation of $\ve{k}$ and $\rax$ and so $\eta_\parallel=0$. b) $\varDelta m=\pm1$-transitions of a OAM-carrying vortex beam with $\sigma=\pm1$: the carrier is forbidden, because one unit of angular momentum has to go into the motion of the ion. In comparison, the axial and both radial sideband modes are excited, because of the non-vanishing projection onto the transversal axes $x$ and $y$ and so $\eta_\perp\neq0$.}
\label{fig:sketch}
\end{figure}

We start by briefly presenting the fundamental equations that determine the possible interaction terms for a structured beam to show that it can transfer its orbital angular momentum to a single trapped ion.  
The quadrupole term of the light-matter interaction Hamiltonian reads: ~\cite{van1994spin, james1997quantum}:
\begin{equation} \label{eq:hamiltonian-principle}
\hat{H}_\mathrm{Q} \sim \sum_{i,j}\hat{q}_i\hat{q}_j \left.\frac{\partial E_j^{(+)}}{\partial q_i}\right\rvert_{\op{Q}_i}  \mathrm{e}^{-\mathrm{i}\omega t} + \mathrm{h.c.},
\end{equation}
where $\hat q_i=\{\op{x},\op{y},\op{z}\}$ are the position operators pertaining to  the valence electron, $q_i$ are the coordinates of the electric field components $E_j$ with angular frequency $\omega$ and the the $\op{Q}_i =\{\op X,\op Y,\op Z\}$ operators act on the ion's center-of-mass coordinates. We consider a beam traveling in the $\ve{e}_z$ direction and a trap coordinates $r_{i}=\{r_\mathrm{R1},r_\mathrm{R2},r_\mathrm{ax}\}$ rotated $45^\circ$ with respect to the $x$ axis, as shown in Fig. \ref{fig:sketch} a) and b).

For the case of a traveling wave, as in the center of a Gaussian beam, the electric field is $\ve{E}_\mathrm{G}^{(+)}\sim E_0(\ve{e}_{x}+\sigma \mathrm{i}\ve{e}_{y})\exp(\mathrm{i}kz)$  with circular polarization $\sigma=\pm1$ and wavenumber $k$, one obtains the usual quadrupole transition selection rules for the valence electron~\cite{schmiegelow2012light,roos2000controlling}. The sideband transitions are governed by the term 
$\op{H}_\mathrm{G}\sim\exp\{\mathrm{i}k\op Z\}$. When expressed in terms of the  eigendirections of ion's motion $\op Z=(\op R_\mathrm{R1}-\op R_\mathrm{R2})/\sqrt{2}$ we see that the only allowed sidebands correspond to the radial components $r_\mathrm{R1}$ and $r_\mathrm{R2}$, while the axial ones $r_\mathrm{ax}=x$ are forbidden, since they are orthogonal to $\ve{k}$, see Fig.~\ref{fig:sketch} a) and c).
The coupling strength of these transitions will be governed by a longitudinal Lamb-Dicke parameter $\eta_{\parallel}=k r_0$, for a quantum harmonic oscillator with $r_0=\sqrt{\hbar/2m\omega_{1,2}}$ being $m$ the mass of the ion and $\omega_{1,2}$ the secular frequencies in the radial directions $r_\mathrm{R1}$ and $r_\mathrm{R1}$. 
Thus, at the center of a Gaussian beam, sideband transitions are driven by the longitudinal electric field gradient, strictly oriented along the propagation direction $\ve{k}$.

The situation is different for a vortex beam carrying OAM, where transverse electric field gradients, perpendicular to the propagation direction, can also mediate sideband excitations, as show in Fig.~\ref{fig:sketch} b) and d). 
At its center, the electric field of a Laguerre-Gaussian beam is: 
$
\ve{E}^{(+)}_\mathrm{LG}\sim E_0\sqrt{2}w_0^{-1} (\ve{e}_{x}+ \mathrm{i}\sigma\ve{e}_{y})\exp(\mathrm{i}kz)(x+\mathrm{i}ly),
$
with 
waist $w_0$, $l=\pm1$ units of orbital angular momentum and $\sigma=\pm1$ units of intrinsic angular momentum.
Upon evaluating the interaction Hamiltonian for this field using Eq.  \ref{eq:hamiltonian-principle}, 
one can sort the result in two terms. The first one contains the coordinates pairs of the transverse directions $\{xx, xy, yx , yy\}$ and accounts for two units of angular momentum of the beam being transferred to its valence electron, as has been demonstrated \cite{schmiegelow2016transfer}. 
Here, the coupling to the external vibrational degrees of motion to the light field is identical as for a Gaussian  beam: only transitions in the direction parallel to $\ve{k}$ are allowed.  

The second term contains crossed longitudinal-transverse coordinate pairs $\{xz, yz\}$ which show a linear dependence on the transverse coordinates:
\begin{equation}
\op{H}_\mathrm{LG}^{\perp}
    \sim
    \exp\{\mathrm{i}k\hat Z\}
    \frac{\sqrt{2}}{w_0}
    \left(
    \hat X+
    \mathrm{i}\hat Y\right).
\end{equation}
This Hamiltonian has a linear dependence on the position operators $\op X=\op{R}_\mathrm{ax}$ and $\op Y=(\op{R}_\mathrm{R1}+\op{R}_\mathrm{R2})/\sqrt{2}$, which result coupling to all the eigendirections of the trap, now allowing for the excitation for the sideband transverse to the beam's propagation direction. Moreover, the carrier term (a constant) is absent, resembling what happens in a standing wave, where either carrier or first sidebands along the direction of the beam can be suppressed by placing a single ion on its nodes or antinodes~\cite{mundt2002coupling}. 
For the vortex beam, the transverse sideband can be excited with a strength at it's center is determined by a \textit{transverse} Lamb-Dicke parameter $\eta_{\perp}=\sqrt{2}x_0/w_0$, which now depends on the beam waist $w_0$ and by the spatial wave packet spread in the axial direction $\rax=x$. 
This describes a quadrupole excitation of the valence electron, changing its magnetic number by $\varDelta m=\pm1$, together with an excitation of the external degree of freedom which, can acquire or give a phonon $\varDelta n_\mathrm{ax}=\pm1$. 
The excitation of a transversal motional degree is enabled by the crosswise spatial gradient of the beam, as illustrated in Fig.~\ref{fig:sketch}~b) and d), while the change in internal angular momentum $\varDelta m$ is determined by its polarization.

\begin{figure*}[htp]
\centering
\includegraphics[width=1\textwidth]{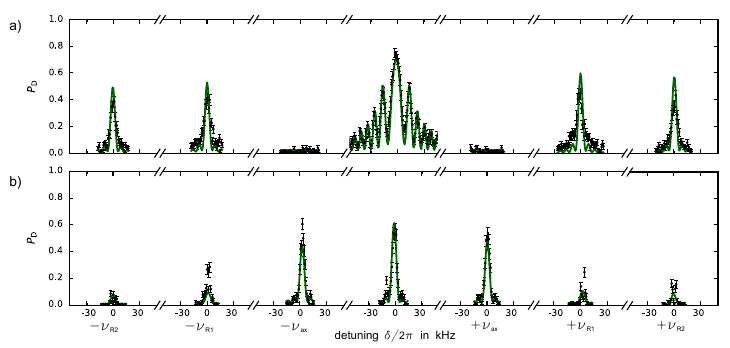}
\caption{Spectrum of the transition $|\mathrm{S}_{\nicefrac{1}{2}}, m_J=-\nicefrac{1}{2}\rangle \leftrightarrow |\mathrm{D}_{\nicefrac{5}{2}}, m_{J'}=-\nicefrac{3}{2}\rangle$ with a) a Gaussian beam and b) a vortex beam. The graph shows the dark state population $P_\mathrm{D}$ for the carrier, axial sidebands ($\delta=\pm2\pi
\nu_\mathrm{ax}$), and radial sidebands ($\delta=\pm2\pi\nu_\mathrm{R1,R2}$). Black dots depict our measurements in frequency steps of $\varDelta \nu=1\,\mathrm{kHz}$, while the green continuous lines map simulations, based on Eq. \ref{eq:rabi_spec}. The selected fix values are chosen from our pulsed laser beam parameters with $P_\mathrm{V}=10\,\upmu\mathrm{W}$, $P_\mathrm{G}=310\,\mathrm{nW}$, $w_\mathrm{G}=2.8\,\upmu\mathrm{m}$, $w_\mathrm{V}=3.3\,\upmu\mathrm{m}$ and $\tau=150\,\upmu\mathrm{s}$, as well as values which base on a Doppler-cooled ion: $\langle n_\mathrm{ax}\rangle = 15$, $\langle n_\mathrm{R1, R2}\rangle = 7$.
}
\label{fig:spectrum}
\end{figure*}

To demonstrate this anomalous momentum transfer, we use an experimental setup where a single $^{40}$Ca$^{+}$ ion is confined in a harmonic potential of quadruple trap mounted in a UHV chamber \cite{wolf2020light}. We use a radio-frequency Paul trap producing a radial harmonic confinement with secular frequencies of $\omega_\mathrm{R1,R2}=2\pi\times(1.70, 2.05)\,\mathrm{MHz}$. The axial confinement is generated by an electrostatic potential, yielding an axial secular frequency of $\omega_\mathrm{ax}\simeq2\pi\times700\;\mathrm{kHz}$. The ion is Doppler cooled in all three trap axes using the dipole transition $4\mathrm{S}_{\nicefrac{1}{2}}\leftrightarrow4\mathrm{P}_{\nicefrac{1}{2}}$ close to $397\,\mathrm{nm}$ with an additional $866\,\mathrm{nm}$ repump laser to form a closed cooling cycle. Single ion imaging is achieved on an EMCCD camera, by collecting the laser-induced fluorescence through an objective with numerical aperture $\mathit{NA}=0.3$, focal distance of 66.9~mm, leading to a diffraction limited image with magnification $M = 15.6(5)$.

The setup is equipped with two $729\,\mathrm{nm}$ laser beams to drive the $4\mathrm{S}_{\nicefrac{1}{2}}\leftrightarrow3\mathrm{D}_{\nicefrac{5}{2}}$ quadrupole transition. A preparation beam, with a Gaussian shape and $\ve{k}$-vector projection onto all three trap axes is used to initialize the ion in the ground state $\ket{\mathrm{S}_{\nicefrac{1}{2}},m_{J}=-\nicefrac{1}{2}}$ and for optional sideband cooling of the axial vibration mode. A second 729~nm beam is used as a probe and can be set to be have either a Gaussian or a  Laguerre-Gaussian transverse mode profile with a holographic pitch-fork pattern. This beam is focused onto the ion though the imaging lens to a beam waist of $w_0\simeq3\,\upmu\mathrm{m}$ (see supplementary material), propagating perpendicular to the trap's axial direction, e.i. along $z$, as seen in Fig.~\ref{fig:sketch}.  Two steering mirrors allow aligning the beam perpendicular to the axial direction with a precision of $<1^\circ $. Moreover, the beam can be scanned across the ion with a range of $\approx 1\,\upmu\mathrm{m}$ in all three spatial directions by using a closed-loop piezo stage. 
A Zeeman splitting of $5\,\mathrm{MHz}$ between the sub-levels in the D$_{\nicefrac{5}{2}}$-state, is generated by magnetic field aligned parallel to the propagation direction of the probe beam. The beam's polarization and frequency are set to drive the $\varDelta m=-1$ transition $\ket{\mathrm{S}_{\nicefrac{1}{2}}, m_{J}=-\nicefrac{1}{2}}\leftrightarrow\ket{\mathrm{D}_{\nicefrac{5}{2}}, m_{J'}=-\nicefrac{3}{2}}$ \cite{schmiegelow2012light}.

We measure the resolved sideband spectra of the Gaussian and the Laguerre-Gaussian with $l=-1$ beam by scanning the probe laser frequency, see Fig.~\ref{fig:spectrum}. An experimental sequence consists of Doppler cooling, spin initialization, pulsed excitation and state dependent fluorescence readout. For the Gaussian beam, Fig.~\ref{fig:spectrum}~a), we observe strong coherent oscillations on the carrier, radial sidebands determined by the longitudinal Lamb-Dicke parameter and absent axial sidebands. 
The experimental data (black) are is good agreement with a model (green) generated by independently measured parameters. 
The spectra pertaining to the Laguerre-Gaussian probe beam shown in Fig. 2 b) show clearly visible axial sidebands, determined by the transverse Lamb-Dicke parameter. 
From the independently measured beam waist and secular frequency, we compute a transverse Lamb-Dicke parameter of  $\eta_{\perp}^\mathrm{theo}=\sqrt{2} x_0/w_0= 0.0057(1)$ (see supplementary material). The difference in strength between axial and radial sidebands is explained by the geometry from the projection of each eigendirection onto the beam's propagation direction and from the wave packet sizes for different mode frequencies. 
As the transversal electric field gradients are smaller by a factor $\sqrt{2}/kw_0$ as compared to the longitudinal ones, the value of the transversal Lamb-Dicke parameter is reduced. We increased the laser power by a factor of $30$, accordingly. 
Note, that also for the vortex beam we observe a residual carrier excitation which is about $2.5\,\%$ as compared to the Gaussian beam shape. 
This is quantitatively explained by accounting for the finite wave packet spatial extension, which
explores an extended region around the center of the beam and then senses an effective
non-vanishing electric field intensity~ \cite{drechsler2021optical}. 
We stress that in all of Fig.~\ref{fig:spectrum} the model was not fitted to the data, but ran with independently measured parameters.

\begin{figure}[tp]
\centering
\includegraphics[width=0.49\textwidth]{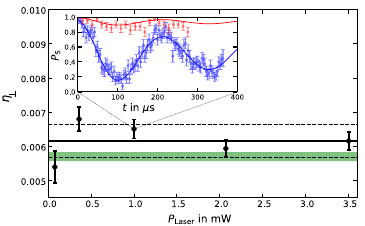}
\caption{Determined mean Lamb-Dicke parameter of $\eta_{\perp}^\mathrm{exp}=0.0062(6)$ (black line) and the $1\sigma$-range (dashed) from fitted $\eta_{\perp}$ for different powers (points). As example, the coherent oscillations of the red and blue transverse sideband for $P_\mathrm{Laser}=1\,\mathrm{mW}$ are shown in the inset. For comparison, the $1\sigma$-range of the theoretically expected value of $ \eta_{\perp}^\mathrm{theo}=0.0057(1)$ is drawn in green with an uncertainty which is mainly due to the experimental error in the determination of the beam waist $w_0$.}
\label{fig:sup_eta}
\end{figure}

Next, we precisely measure the transverse Lamb-Dicke parameter $\eta^\mathrm{exp}_{\perp}$ for the Laguerre-Gaussian beam by analyzing Rabi oscillations on the axial red and blue sidebands, see Fig.~\ref{fig:sup_eta}. The ion was cooled near to the ground state of axial motion to a mean phonon number $\langle n_\mathrm{ax}\rangle =0.19(10)$ to obtain good contrast of the sideband Rabi oscillations. From the data of the red and blue sidebands, we extract $\eta^\mathrm{exp}_{\perp}$, using the independently measured Rabi frequency and mean phonon number as fixed parameters, by a simultaneous fit to both curves.  Measurements were taken for five different beam powers between 78~$\upmu$W and 3.5~mW, in order to exclude any dependence on the beam power. By averaging, we obtain a value of $\eta^\mathrm{exp}_{\perp}=0.0062(5)$ being in agreement with the theoretical value of $\eta^{\mathrm{theo}}_{\perp}=0.0057(1)$, all shown in Fig.~\ref{fig:sup_eta}.

\begin{figure}[tp]
\centering
\includegraphics[width=0.49\textwidth]{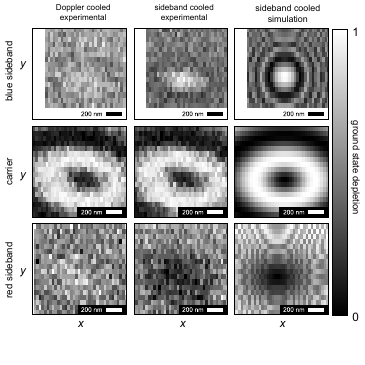}
\caption{Profile measurements in $x$ and $y$ directions by scanning the beam for carrier (car), red sideband and blue sideband (sb) of the $|\mathrm{S}_{\nicefrac{1}{2}}, m_{J}=-\nicefrac{1}{2}\rangle \leftrightarrow |\mathrm{D}_{\nicefrac{5}{2}}, m_{J'}=-\nicefrac{3}{2}\rangle$ transition with $P_\mathrm{car}=45\,\upmu\mathrm{W}$ and $P_\mathrm{sb}=2.8\,\mathrm{mW}$. Situation is shown for Doppler-cooled (left column) and a sideband-cooled ion close to the ground state (middle column). The pulse duration of $\tau=55\,\upmu\mathrm{s}$ corresponds to the blue sideband's $\uppi$-time of the sub-cooled ion. Right column shows numerical simulations of the expected profiles for the sideband-cooled case.}
\label{fig:2dplots}
\end{figure}

Using the ion as a probe, we explore the structural dependence of the sideband and the carrier transitions excited by the Laguerre-Gaussian beam. We vary the wave packet size using an ion either Doppler cooled with $\langle n_\mathrm{ax}\rangle \approx15$ or cooled close to its motional ground state in the axial direction, as shown in \hyperref[fig:2]{Fig. \ref{fig:2dplots}}. We scan the beam over the ion in the transverse $x$ and $y$ directions, with a 
step sizes of 32~nm ($x$) and 64~nm ($y$)
and record the excitation probability for an effective pulse area of about $\uppi$  for ground-state cooled axial blue sideband, corresponding to a rectangular pulse with a duration of 55~$\mu$s.
The carrier shows a ring-structure, due to the increase of the electric field
amplitude radially from the beam's center~\cite{drechsler2021optical}. Conversely, the sidebands display increased excitation at the center of the vortex beam, due to the maximum of the transversal gradient. This is seen for all cases except for the red sideband of the sub-Doppler cooled ion, where excitation is suppressed by the fact that the oscillator state is near $\ket{n_\mathrm{ax}=0}$. The maximum excitation can be achieved on the transversal blue sideband in the dark center of the vortex beam. Moreover, we note all the sidebands display a halo structure, which is caused by the off-resonant excitation of the carrier, as we verify by numerical simulation of the expected profiles, see right column in  Fig.~\ref{fig:2dplots}.

We demonstrated how the spatial structure of a shaped laser beam can import momentum on the center-of-mass motion of atomic particles and even drive coherent ion-light interaction using the vortex light field. We believe this technique might find various applications. For example, in quantum computing with linear ion chains, where radial addressing beams, could be used to drive gates  
mediated by axial modes, which are less subject to spectral crowding as compared to radial modes and which are amenable to sympathetic cooling~\cite{hilder2022fault,pogorelov2021compact,debnath2016demonstration}. Also, the reduced carrier might help exploring sideband cooling limits for trapped ions or help sensing and manipulating local phonon occupation numbers for planar ion crystals intended for quantum simulations~ \cite{espinoza2021engineering,monroe2021programmable,bond2022effect}. Finally, we note that, by using dipolar transitions, one could observe torques on single ions or ion crystals by resonant and off-resonant forces generated by the transverse field gradients of vortex beams.\\

We acknowledge the Deutsche Forschungsgemeinschaft within the TRR 306 (QuCoLiMa, Project-ID 429529648) (Germany), the  Alexander von
Humboldt-Stiftung/Foundation (Germany), and the Agencia Nacional de Promoción de la Investigación, el Desarrollo Tecnológico y la Innovación (Argentina), for financial support. We thank U. Poschinger and A. Trimeche for careful reading and comments on the manuscript.

\bibliography{bibliography.bib}

\pagebreak

\section{Supplementary Material}

\paragraph{\textbf{Theoretical framework}}
We present here the formalism to describe carrier and sidebands excitations, shown in \hyperref[fig:]{Fig. \ref{fig:spectrum}} and \hyperref[fig:]{\ref{fig:2dplots}}, which are driven by spatially structured light, addressing electronic and motional degrees of freedom of a single trapped ion. In the context of semi-classical theory with quantized matter and classical fields, and by following the derivation of van Enk and James~\cite{van1994spin, james1997quantum}, the quadrupole term of light-matter interaction $\hat{H}_\mathrm{Q}$ is:
\begin{equation} \label{interaction hamiltonian}
\hat{H}_\mathrm{Q} = \sum_{i,j}\hat{q}_i\hat{q}_j \partial_i E_j^{(+)} \mathrm{e}^{-\mathrm{i}\omega t} + \mathrm{h.c.},
\end{equation}

where we neglect two higher orders terms, one accounting for interactions between two electrons and the nucleus and the other being the Röntgen term. The usual rotating wave approximation is then applied and the positive-frequency part $\hat{H}_\mathrm{Q}^{(+)}$, which only contributes to matrix elements, is decomposed by means of spherical tensor operators $\hat{T}_{\varDelta m}^{\varDelta l}$. For quadrupole transitions, the ion acquires a total angular momentum $\varDelta l=2$, with possible projections on the quantization axis being $\varDelta m=0,\pm1,\pm2$. Within the paraxial approximation, with no longitudinal electric field component, $\hat{H}_\mathrm{Q}^{(+)}$ reduces to: 

\begin{equation} \label{Paraxial-Hamiltonian}
\begin{split}
\hat{H}_\mathrm{Q}^{(+)}  = |\hat{\ve{Q}}|^2\frac{2\pi}{3}&[\hat{T}^{2}_{2}\big( \partial_x E_x - \partial_y E_y - \mathrm{i} \partial_x E_y - \mathrm{i} \partial_y E_x \big)  + \\
& +\hat{T}^{2}_{1}\big(\partial_z E_x - \mathrm{i} \partial_z E_y \big)  + \\
& +\hat{T}^{2}_{0}\big( \frac{2}{\sqrt{6}}(\partial_x E_x + \partial_y E_y ) \big)  + \\
& +\hat{T}^{2}_{-1}\big( \partial_z E_x + \mathrm{i} \partial_z E_y \big)  + \\
& +\hat{T}^{2}_{-2}\big( \partial_x E_x - \partial_y E_y + \mathrm{i} \partial_x E_y + \mathrm{i} \partial_y E_x \big)],
\end{split}
\end{equation}
which, applied to light fields, allow us to compare the various quadrupole excitations experienced by the ion. Near the center, the electric field for a Gaussian and a vortex beam with topological charge $l=\pm 1$, are respectively:

\begin{equation}
   \ve{E}_\mathrm{G}(x,y,z) \sim \frac{E_0}{\sqrt{\pi}}\mathrm{e}^{\mathrm{i} k z}\frac{(\ve{e}_x + \mathrm{i}\sigma\ve{e}_y )}{\sqrt{2}}
\end{equation}

\begin{equation}
   \ve{E}_\mathrm{LG}(x,y,z) \sim \frac{E_0}{\sqrt{\pi}}\frac{(x\pm\mathrm{i}y)}{w_0}\mathrm{e}^{\mathrm{i} k z}(\ve{e}_x + \mathrm{i}\sigma\ve{e}_y ),
\end{equation}
where $E_0$ depends on the power, $w_0$ is the beam waist and $\sigma=\pm1$ is the circular polarization. By evaluating the electric field partial derivatives up to first order in the coordinates $x$, $y$ and $z$, and then promoting them to quantum operators, expressed in terms of annihilation and creation operators $\hat{a}_i$ and $\hat{a}_i^\dagger$ for the harmonically trapped ion, we finally have:

\begin{eqnarray} \label{Gaussian-Hamiltonian}
    \op{H}_\mathrm{G} 
    &\sim&
    |\hat{\ve{Q}}|^2\frac{\sqrt{8\pi}}{3}\mathrm{i} k E_0  \op{T}_{\sigma}^{2}
    \exp\left(\mathrm{i}kz_0(\op{a}_z+\op{a}_z^\dagger)\right),
\end{eqnarray}

in the Gaussian case, which only allows  $\varDelta m=\pm1$ transitions with longitudinal red and blue sidebands mediated by $\op{a}_z$ and $\op{a}_z^\dagger$ respectively. While for twisted light, we have a sum of two different terms. The first is:

\begin{eqnarray}    \label{Vortex-Hamiltonian-perpendicular}
    \op{H}_{\mathrm{LG},\perp} 
    &\sim&
   |\hat{\ve{Q}}|^2\frac{\sqrt{8\pi}}{3}\mathrm{i} k E_0 \,  \op{T}_{\sigma}^{2}\,
    \exp\left(\mathrm{i}kz_0(\op{a}_z+\op{a}_z^\dagger)\right)
    \times\nonumber\\&&
    \left(\frac{\sqrt{2} x_0}{w_0}(\op{a}_x+\op{a}_x^\dagger)\pm
    \mathrm{i}\frac{\sqrt{2} y_0}{w_0}(\op{a}_y+\op{a}_y^\dagger)\right),
\end{eqnarray}

and mediates $\varDelta m=\pm 1$ transitions, nicely predicting transverse sidebands with no carrier excitation, firstly reported in \hyperref[fig:]{Fig. \ref{fig:spectrum}}. The second term is:

\begin{eqnarray}     \label{Vortex-Hamiltonian-parallel}
    \hat{H}_{\mathrm{LG},\parallel}  
     &\sim&
     |\hat{\ve{Q}}|^2\frac{\sqrt{8\pi}}{3}E_0 \, \frac{\sqrt{2}}{w_0} \exp\left(\mathrm{i}kz_0(\op{a}_z+\op{a}_z^\dagger)\right)
    \times\nonumber\\&&
    (\delta_{\pm,1}\op{T}_{2}^{2}(1+\sigma)+\frac{1}{\sqrt{6}}\op{T}_{0}^{2}(1\mp\sigma)+\\&&
    +\delta_{\pm,-1}\op{T}_{-2}^{2}(1-\sigma)),\nonumber
\end{eqnarray}

which drives, in the vortex beam singularity, different $\varDelta m=0,\pm2$ carrier excitations and their corresponding longitudinal sidebands. Referring to $\varDelta m=-1$ transition mediated by $\op{H}_{\mathrm{G}}$ and $\op{H}_{\mathrm{LG},\perp}$, we compare Eqs. \ref{Gaussian-Hamiltonian} and \ref{Vortex-Hamiltonian-perpendicular} to define a new transverse Lamb-Dicke parameter $\eta_\perp$:

\begin{align} 
    &\eta_{{i},\perp}  = \ve{n}_{\perp}\cdot \ve{e}_{{i}} \frac{\sqrt{2}}{w_0} \sqrt{\frac{\hbar}{2m\omega_{i}}},
    \label{eq:perp-Lamb-Dicke}
\end{align}

with the usual longitudinal Lamb-Dicke parameter being $\eta_{{i},\parallel}  =  \ve{k}\cdot \ve{e}_{{i}} \sqrt{\frac{\hbar}{2m\omega_{i}}}$. Now, even if Eq. \ref{eq:perp-Lamb-Dicke} is strictly valid in the vortex beam singularity, $\eta_\perp$ can be evaluated for any light-field position, the result being essentially a space-modulation with the electric field transverse gradient. The arising spatial structure for transverse sidebands has been carefully measured in Fig. \ref{fig:2dplots} to be in good agreement with our model. Finally, we stress that $\eta_\perp$, here derived for twisted-light, is more properly related to electric field transverse gradients, then becoming relevant whenever structured light features this property.

\begin{figure}[htp]
\centering
\includegraphics[width=0.49\textwidth]{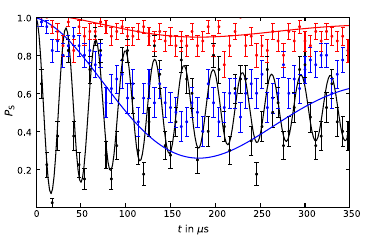}
\caption{For the sideband cooled ion, Rabi oscillations for the $|\mathrm{S}_{\nicefrac{1}{2}}, m_{J}=\nicefrac{1}{2}, n_\mathrm{ax}\rangle \leftrightarrow |\mathrm{D}_{\nicefrac{5}{2}}, m_{J'}=\nicefrac{5}{2}, n_\mathrm{ax}\rangle$ carrier (black), $| n_\mathrm{ax}\rangle \leftrightarrow |n_\mathrm{ax}-1\rangle$ (red) sideband and  $| n_\mathrm{ax}\rangle \leftrightarrow |n_\mathrm{ax}+1\rangle$ (blue) sideband are recorded. By fitting the data with Eq. \ref{eq:sub_rabi}, a mean phonon number $\langle n_\mathrm{ax}\rangle=0.19(10)$ in the axial direction, is obtained.
}
\label{fig:sup_temp}
\end{figure}

\paragraph{\textbf{Measured transverse Lamb-Dicke parameter}}
We describe here, in some details, the experimental determination of $\eta_{\perp}^\mathrm{exp}$, which can therefore be compared to the predicted value $\eta_{\perp}^\mathrm{theo}$ from Eq. \ref{eq:perp-Lamb-Dicke}. Our measurement essentially relies on the Rabi frequency evaluation for the transverse blue sideband as coherently excited with the trapped ion being in its motional ground state at the center of twisted-light. The requirement of motional ground state, even if not necessary, is highly desirable. Indeed, the presence of many phonons unavoidably leads to dephased Rabi oscillations, then resulting in way less precise $\eta_{\perp}^\mathrm{exp}$ determination. In our setup, the Doppler cooled ion is characterized by a mean phonon number $\langle n_\mathrm{ax}^{DC}\rangle \approx 15$ in the axial direction, then additional cooling mechanisms are needed. We then perform sideband cooling on the $|\mathrm{S}_{\nicefrac{1}{2}}, m_{J}=\nicefrac{1}{2}, n_\mathrm{ax}\rangle \leftrightarrow |\mathrm{D}_{\nicefrac{5}{2}}, m_{J'}=\nicefrac{5}{2}, n_\mathrm{ax}-1\rangle$ by using a laser beam, steered at $45^\circ$ to the axial direction, which has a projection of its $\ve{k}_{729}$-vector on this motional mode. As shown in Fig. \ref{fig:sup_temp}, the sideband cooled ion is close to its motional ground state, with a low mean phonon number of $\langle n_\mathrm{ax}\rangle=0.19(10)$.

\begin{figure}[htp]
\centering
\includegraphics[width=0.49\textwidth]{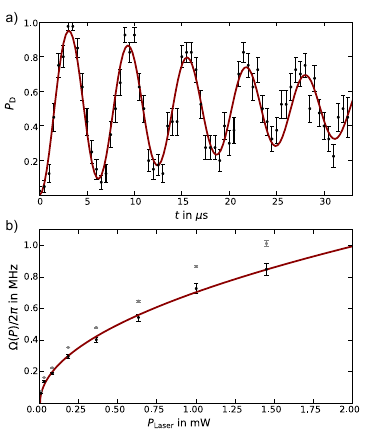}
\caption{a) Pulse width scan of the carrier transition $|\mathrm{S}_{\nicefrac{1}{2}}, m_{J}=-\nicefrac{1}{2}\rangle \leftrightarrow |\mathrm{D}_{\nicefrac{5}{2}}, m_{J'}=-\nicefrac{3}{2}\rangle$ by using a Gaussian beam with $P_\mathrm{Laser}=26\,\upmu\mathrm{W}$. Fit function (Eq. \ref{eq:sub_rabi}) provides the Rabi frequency $\Omega$ of the coherent oscillation. b) Power-dependend Rabi frequency determined by a square-root fit function with $\Omega(P)=2\pi\cdot0.704(8)\,\mathrm{MHz}\cdot\sqrt{\frac{P}{\mathrm{mW}}}$. Grey points indicate measured data, black points are the corrected values due to the waist shift.}
\label{fig:sup_rabi_cali}
\end{figure}

The value $\langle n_\mathrm{ax}\rangle=0.19(10)$ is established by fitting the data shown in Fig. \ref{fig:sup_temp} with the general Rabi cycling model for the carrier, red and blue sideband population:

\begin{align}
    P_\mathrm{D}(t)&=\frac{1}{2}\left(1-\sum_{n=0}^{\infty} \rho_n\cos(\Omega_{n,m} t) \mathrm{e}^{-\Gamma t}\right),
\label{eq:sub_rabi}
\end{align}

where $\Gamma$ describes the overall decoherence, mainly coming from magnetic field fluctuations, $\rho_n=\frac{\langle n\rangle^n}{(\langle n\rangle+1)^{n+1}}$ is the thermal phonon distribution and $\Omega_{n,m}$ are the different Rabi frequencies, respectively being $\Omega_{n,n}=(1-\eta_{\parallel}^2n)\Omega_0$ for the carrier, $\Omega_{n-1,n}=\eta_{\parallel}\sqrt{n}\Omega_0$ for the red sideband and $\Omega_{n+1,n}=\eta_{\parallel}\sqrt{n+1}\Omega_0$ for the blue sideband, where $n$ is the number of phonons and $\eta_{\parallel}=\ve{e}_x\ve{k}_\mathrm{729}\sqrt{\frac{\hbar}{2m\omega_\mathrm{ax}}}=0.0819(1)$ is the Lamb-Dicke parameter for the sideband cooling geometry. Now, the value of $\langle n_\mathrm{ax}\rangle$ can be used a fixed parameter when evaluating $\eta_{\perp}^\mathrm{exp}$ from the Rabi frequency of transverse blue sideband. In order of doing that, we use the model described at Eq. \ref{eq:sub_rabi} restricted to the blue sideband, with the minor substitution $\eta_{\parallel}\rightarrow\eta_{\perp}$, with the frequency $\Omega_{n+1,n}=\eta_{\perp}\sqrt{n+1}\Omega_0$ being then used to evaluate $\eta_{\perp}^\mathrm{exp}$, once $\Omega_0$ is known. This is commonly determined as the Rabi frequency for the carrier transition of a ground state cooled ion. Since no carrier is excited in the vortex beam singularity, we do it after switching to the Gaussian mode. As shown in Fig. \ref{fig:sup_rabi_cali}, we record the Rabi dynamics for different laser beam powers between $5.3\,\upmu\mathrm{W}$ and $1.45\,\mathrm{mW}$, finally determining $\Omega_{0,\mathrm{P}}=2\pi\cdot0.704(8)\,\mathrm{MHz}\cdot\sqrt{\mathrm{mW}}^{-1}$, which is then used for the evaluation of $\Omega_0$ at any desired power.

Given $\langle n_\mathrm{ax}\rangle$ and $\Omega_{0,\mathrm{P}}$, we switch the laser back to the vortex mode and we measure the Rabi oscillations of the transverse sidebands for different powers between $78\,\upmu\mathrm{W}$ and $3.5\,\mathrm{mW}$. As shown in Fig. \ref{fig:sup_eta}, we finally obtain the average value $\eta_{\perp}^\mathrm{exp}=0.0062(6)$.

\paragraph{\textbf{Experiment to theory comparison}}

\begin{figure}[htp]
\centering
\includegraphics[width=0.49\textwidth]{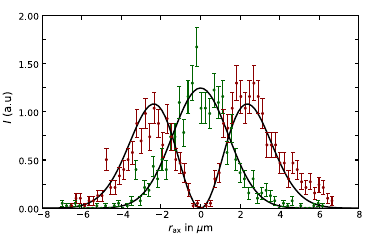}
\caption{Waist size measurement for focused Gaussian beam (green) and focused vortex beam (red). The laser power in both cases were below $5\,\upmu\mathrm{W}$. Continuous lines are the fit for the excited population. The beam waist are slightly different with $w_\mathrm{G}=2.8(1)\,\upmu\mathrm{m}$ and $w_\mathrm{V}=3.34(7)\,\upmu\mathrm{m}$. This difference systematically occurs when we switch between the two types of modes. No focal plane displacement is present and the focusing objective is responsible for this difference, as the beam waist measured before it, show no difference. }
\label{fig:sup_waist}
\end{figure}

From Eq. \ref{eq:perp-Lamb-Dicke}, the transverse Lamb-Dicke parameter $\eta_{\perp}^\mathrm{theo}=\frac{\sqrt{2}}{w_\mathrm{V}} \sqrt{\frac{\hbar}{2m\omega_\mathrm{ax}}}$ depends on the axial trap frequency $\omega_\mathrm{ax}$ and on the vortex beam waist $w_\mathrm{V}$, two tunable values that have to be measured. The trap frequencies were precisely determined by \textit{tickling} the ion with a variable RF signal fed at one of the ion trap's electrodes and observing its motion on the fluorescence image. They were moreover double checked by independently varying DC and RF voltages, and verifying the expected shifts. The axial trap frequency is then measured to be $\omega_\mathrm{ax}\simeq2\pi\times700(1)\;\mathrm{kHz}$. The vortex beam waist, instead, has been measured by moving the ion perpendicularly through the beam, along the trap axis, while driving its resonant transition $|\mathrm{S}_{\nicefrac{1}{2}}, m_{J}=-\nicefrac{1}{2}\rangle \leftrightarrow |\mathrm{D}_{\nicefrac{5}{2}}, m_{J'}=-\nicefrac{3}{2}\rangle$ and measuring its excited state population, as shown in Fig. \ref{fig:sup_waist}.

In order to get the result in suitable units, we previously calibrate the spatial shift $s=2.20(1)\,\nicefrac{\upmu\mathrm{m}}{\mathrm{V}}$ due to
applied voltages, by moving a two-ion crystal across the
vortex beam and then equating the known theoretical
distance between the two ions to the voltage difference
where each single ion overlaps the vortex dark spot. We stress here that this procedure is consistent, since our Paul trap, with eleven individually addressable DC segments, uses the middle electrode to control the axial confining potential, and uses the second outermost electrode to contemporarily tune the ion's axial position, without changing the trap frequencies. We finally obtain the vortex beam waist $w_\mathrm{V}=3.34(7)\,\upmu\mathrm{m}$, which together with the measured axial trap frequency is used to calculate the theoretical transverse Lamb-Dicke parameter. It is worth to underline here the full agreement, within the $1\sigma$-range, between the previously determined experimental value and the theoretical prediction:

\begin{eqnarray}
    &\eta_{\perp}^\mathrm{exp}=0.0062(6), \nonumber 
    \\&\eta_{\perp}^\mathrm{theo}=0.00570(12).
\end{eqnarray}

\paragraph{\textbf{Spectrum analysis}}
It is worth underlying that the agreement between the theory and our measurements goes far beyond the determination of the single transverse Lamb-Dicke parameter in the axial direction. Indeed, as shown in Fig. \ref{fig:spectrum} and Fig. \ref{fig:2dplots}, we can reproduce, respectively, the whole spectra involving transverse and longitudinal sidebands, and their spatial dependency within the twisted-light singularity. Our model consists in the following sum over the phonon distribution $\rho_n$ to calculate the space-dependent dark state population $P_\mathrm{D}(x,y)$ for a given detuning $\delta$ between the laser frequency $\nu_\mathrm{L}$ and the transition frequency $\nu_0$:

\begin{align}
    P_\mathrm{D}(x,y)&=\sum_{n=0}^{\infty} \rho_n\frac{\Omega_{n,\nu_0}^2}{\Omega_{n,\nu_0}^2+\delta^2}\sin^2\left(\frac{\tau}{2}\sqrt{\Omega_{n,\nu_0}^2+\delta^2}\right) \mathrm{e}^{-\Gamma \tau},
\label{eq:rabi_spec}
\end{align}

where the spatial dependence is contained in the Rabi frequency $\Omega_{n,\nu_0}(x,y)$ for the particular transition, i.e. the carrier or the sidebands, and depending on the specific beam profile, here Gaussian or vortex.

\begin{figure}[t]
\centering
\includegraphics[width=0.49\textwidth]{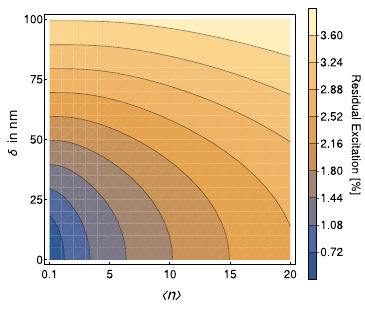}
\caption{Residual carrier excitation depending on on the ion's displacement $\delta$ from the vortex beam center and on the wave packet size, increasing with the mean phonon number $\langle n \rangle$. Both parameters affect the effectively sensed electric field intensity. A previously determined, unavoidable thermo-mechanical drift of the setup, of $\delta=50\,\mathrm{nm}$ (over the duration of a typical measurement), joined with the value of $\langle{n}\rangle =15$ phonons, gives rise to a residual excitation of 2.5 $\%$ in excellent agreement with the measurement shown in \hyperref[fig:]{Fig. \ref{fig:spectrum}}.}
\label{fig:residual-excitation}
\end{figure} 

In particular, longitudinal and transverse sidebands, respectively rely on the different $\eta_{\parallel}$ and $\eta_{\perp}$ Lamb-Dicke parameters. For our setup geometry, they are predicted to be:

\begin{tabular}{ |p{2.5cm}|p{2.5cm}|p{2.5cm}| }
  \hline
 &Gaussian $\eta_{\parallel}$   &vortex $\eta_{\perp}$\\
 \hline
axial ax& $0.0$   & $0.0057$\\
radial R1& $0.053$   & $0.0026$\\
radial R2& $0.048$   & $0.0024$\\
 \hline
\end{tabular}

By using these values with Eq. \ref{eq:rabi_spec}, the spectrum shown in \hyperref[fig:]{Fig. \ref{fig:spectrum}} is accurately reproduced. However, special attention is needed when dealing with the carrier excitation driven within the vortex beam, where it should be fully suppressed as the electric field intensity perfectly vanishes, see Eq. \ref{Vortex-Hamiltonian-perpendicular}. This apparent discrepancy is solved by accounting for the finite size of the ion wave packet, which explores an extended region and then senses an effective non-vanishing electric field intensity. Phonon numbers of $\langle n_\mathrm{ax}\rangle = 15$ and $\langle n_\mathrm{R1, R2}\rangle = 7$ were assumed, which corresponds to wave-packet sizes of $\sim 75\,\mathrm{nm}$ in axial direction and $\sim30\,\mathrm{nm}$ in radial directions. 
By integrating the thermal wave packet, characterized by different average phonon number and centered with a certain displacement away from the vortex beam center, we quantified the residual excitation to be about $2.5\,\%$, see \hyperref[fig:]{Fig. \ref{fig:residual-excitation}} and by using this value we found an excellent agreement with the measured residual carrier excitation amplitude driven by the Vortex beam.

\end{document}